\def\revtex{2}

\ifnum\revtex>0 
\documentstyle[aps,prb,preprint,epsfig,amsmath,amssymb,verbatim]{revtex}

\else
\documentclass[10pt]{article}
\usepackage{epsfig}
\usepackage[]{graphicx}
\DeclareGraphicsExtensions{.eps}
\usepackage{times}
\usepackage{a4wide}
\usepackage{xtdeqnra}
\usepackage{amsmath,amssymb}
\usepackage{ltxtable}
\usepackage{setspace}
\usepackage{verbatim,lscape}
\usepackage{cite}
\usepackage{apthm,multicol,paralist}

\fi

\DeclareGraphicsExtensions{.eps}
\graphicspath{{Figures/}}

\newcommand{\lra}[1]{\langle #1 \rangle }
\newcommand{\mb}[1]{\mathbf{#1}}

\begin{document}

\setcounter{totalnumber}{10}

\title{Langevin PDF simulation of particle deposition in a turbulent pipe flow}
\author{Sergio Chibbaro$^{1}$, Jean-Pierre Minier$^{2}$ \\
$^{1}$  Dept. of Mechanical Engineering, University of ``Tor Vergata'', \\
Viale Politecnico, Rome, Italy \\
chibbaro@iac.cnr.it \\
$^{2}$Electricit\'e de France, Div. R\&D, MFEE, \\
6 Quai Watier, 78400 Chatou, France \\
E-mail: Jean-Pierre.Minier@edf.fr\vspace{2mm} }
\maketitle

\begin{abstract}
The paper deals with the description of particle deposition on walls
from a turbulent flow over a large range of particle diameter,
using a Langevin PDF model. The first aim of the work is to test how 
the present Langevin model is able to describe this phenomenon and to 
outline the physical aspects which play a major role in particle deposition.
The general features and characteristics of the present stochastic model
are first recalled. Then, results obtained with the standard form of the 
model are presented along with an analysis which has been carried out to check 
the sensitivity of the predictions on different mean fluid quantities.
These results show that the physical representation of the near-wall physics
has to be improved and that, in particular, one possible route is to introduce
specific features related to the near-wall coherent structures. In the following,
we propose a simple phenomenological model that introduces some of the effects 
due to the presence of turbulent coherent structures on particles in a thin layer
close to the wall. The results obtained with this phenomenological model 
are in good agreement with experimental evidence and this suggests to pursue in
that direction, towards the development of more general and rigorous stochastic 
models that provide a link between a geometrical description of turbulent flow 
and a statistical one.

\end{abstract}

\section{Introduction}

Particle deposition from a turbulent flow on walls is an important phenomenon
which is observed in many engineering applications, 
for example thermal and nuclear systems, 
cyclone separators, spray cooling and which is also
present in various environmental situations.
Given the large number of possible applications, 
a lot of interest has been devoted to this subject
and many  studies have been carried out in the last  decades.

Different experiments have been conducted to observe deposition
in turbulent flows. In most of them, attention is focused on the
deposition velocity~\cite{Liu_74,McC_77} which is defined as 
$k_p = m_p/\bar{C}$, 
where  $m_p$ is the mass flux and $\bar{C}$ is the bulk mean particle
concentration. This deposition rate, often presented as 
the dimensionless deposition velocity $k_p/u^*$, is a function
of  the dimensionless particle relaxation time,
$\tau_p^+$ defined as
\begin{equation}
\tau_p^+ = S^+\frac{u^*}{U_{p0}} = \frac{d_p^2 \rho_p U_{p0} u^*}{18\mu_f\nu_f}
\frac{u^*}{U_{p0}} = \frac{d_p^2 \rho_f^2 u^{*2}}{18\mu_f^2}\frac{\rho_p}{\rho_f}
\label{taup}
\end{equation}
where $S^+$ is the dimensionless stopping distance,
$U_{p0}$ is the particle initial velocity and $u^*$
the friction velocity.
In this work, $u^*$ has been computed with the Blasius formula,
$u^* = [0.03955Re^{0.25}]^{0.5}U_m$, 
with $U_m$ the bulk mean velocity. 
The deposition velocity is indeed the key point in many engineering applications
where the interest is to obtain the curve that gives 
$k_p/u^*$ as a function of $\tau_p^+$, that is as a function of the
particle diameter. 
Recently, several experimental studies and DNS studies 
of particle deposition have been
presented~\cite{Eat_94}$^-$~\cite{Van_98} and have improved the
understanding of the physical mechanisms at play. 
In particular, much information has been obtained about
the dynamical structures of wall-bounded flows, such as
the coherent structures which manifest themselves in the near-wall 
region.
It is largely accepted that particle transfer in the wall region and also 
deposition onto walls are processes dominated by near-wall turbulent coherent 
structures (sweeps and ejections), which are instantaneous realizations of 
the Reynolds stresses, and that particles tend to remain trapped along the streaks 
when in the viscous-layer~\cite{Eat_94,Kaf_95a,Kaf_95b,Rou_01,Mar_02}.
However, the importance of these mechanisms for particle deposition depends on 
particle inertia. In a somewhat crude picture, light particles follow closely 
sweeps and ejections and their motion towards the wall appears to be very 
well correlated with turbulent structures. Therefore, they are found to deposit 
mainly with negligible wall-normal velocities and large near-wall residence time.
This mechanism of deposition has been called {\it diffusional}~\cite{Nar_03}.
On the contrary, heavy particles  are not so well correlated with turbulent 
structures and their motion is less influenced by them in the near-wall region.
Therefore, heavy particles deposit with large wall-normal velocities and small 
near-wall residence time, that is  by the so-called {\it free-flight} 
mechanism ~\cite{Fri_57,Nar_03}.

Considering the engineering importance of the subject, models that reach
acceptable compromise between simplicity and accuracy are  needed.
While DNS calculations may be regarded as numerical experiments and 
give access to the complete picture, they remain limited to simple
geometries and low-Reynolds number flows. Therefore, a statistical
approach is still necessary to describe the motion of particles in a turbulent
flow. Within this framework, and since the objective is to simulate the entire 
curve of the deposition velocity for a whole range of particle inertia or diameter,
a Lagrangian approach appears appropriate. Indeed, in this approach, the
trajectories of individual particles are tracked and polydispersion
is treated without approximation. The influence of the underlying turbulent
fluid is represented, in the particle equation of motion, by stochastic models. 
Many of the Lagrangian models proposed today belong to the class of the so-called 
{\it random-walk} models~\cite{McI_92,Mat_00,Kro_00}, which define the velocity 
as the sum of the local mean fluid velocity and a random fluctuating velocity 
sampled from a Gaussian distribution. Unfortunately, these models can suffer 
from problems of consistency, in particular the so-called spurious drift effect. 
This is important for particle deposition, since one has to simulate the behavior 
of very small particles which nearly represent fluid tracers. 
In the present paper, we use a Langevin model~\cite{Min_04} in which the velocity 
of the fluid seen by particles is simulated by a diffusion stochastic process. 
This model is consistent in the tracer limit by construction, and is thus free of 
spurious drifts~\cite{Pop_87,McI_92}. Furthermore, the model is formulated in 
terms of instantaneous variables which allows a direct introduction of external 
information provided by fundamental studies (DNS, experiments).

The present numerical Langevin model is applied to a case of particle deposition
in a turbulent pipe-flow. A first purpose is to analyze how the present form of the 
Langevin model performs for particle deposition. A second purpose is to bring out 
the modeling points that are important in this situation so that directions
of improvement are clearly indicated. In particular, a new phenomenological model 
which takes into account some aspects due to the presence of near-wall instantaneous 
coherent structures will be proposed. In this way, we propose a first link between 
a statistical model, such as  the present Langevin model, and some geometrical 
features recently found out by DNS analysis \cite{Mar_02,Mar_03,Nar_03,Pic_05}.
The goal of the work is therefore to propose simple and phenomenological models 
and, also, to indicate whether introducing geometrical features in a Lagrangian 
stochastic approach can be useful for particle deposition simulations.
This approach has some analogy with the analysis carried out by Pope and Yeung some 
years ago for the single-phase fluid stochastic modeling \cite{Yeu_89}.

The paper is  divided as follows. In section~\ref{sec:theory}, we present the Langevin 
model that will be used throughout the work. In section~\ref{sec:results}, we present 
the test-case that will be studied. Results obtained with the standard form of the 
PDF model are discussed and a new phenomenological model for the effect of near-wall 
structures is proposed. Finally, conclusions are proposed.

\section{Langevin Model}
\label{sec:theory}

In this section we recall briefly the theoretical background of turbulent 
two-phase flows and we present the Langevin stochastic model which be will 
referred to as the standard model and which will be used in following 
numerical investigations. The modeling starting point is the exact equations 
of motion. Since two different phases are present, the continuous one and the 
discrete dispersed one, the complete problem is described by two sets of equations.
The continuous phase is described by the Navier-Stokes equations:
\begin{subequations}
\label{fluid: exact field eqs.}
\begin{align}
&\frac{\partial U_{f,j}}{\partial x_j}=0, \\
&\frac{\partial U_{f,i}}{\partial t}+U_{f,j}\frac{\partial U_{f,i}}{\partial x_j}
= -\frac{1}{\rho_f}\frac{\partial P}{\partial x_i} + 
   \nu \frac{\partial^2 U_{f,i}}{\partial x_j^2},
\end{align}
\end{subequations}
while the discrete particle equations in the limit $\rho_p \gg \rho_f$ are \cite{Max_83,Gat_83}
\begin{subequations} 
\label{exact particle eqns}
\begin{align} 
& \frac{d\mb{x}_p}{dt} = \mb{U}_p, \\
& \frac{d\mb{U}_p}{dt} = \frac{1}{\tau_{p}}(\mb{U}_{s}-\mb{U}_{p})
\;  + \mb{g},
\end{align}
\end{subequations}
where $\mb{U}_{s}=\mb{U}(\mb{x}_{p}(t),t)$ is the fluid velocity seen,
\textit{i.e.} the fluid velocity sampled along the particle trajectory
$\mb{x}_{p}(t)$, not to be confused with the fluid velocity $\mb{U}_{f}=
\mb{U}(\mb{x}_{f}(t),t)$ denoted with the subscript $f$. The particle
relaxation time is defined as
\begin{equation} \label{definition taup}
\tau_{p}=\frac{\rho_p}{\rho_f}\frac{4d_p}{3 C_D |\mb{U_r}|},
\end{equation}
where the local instantaneous relative velocity is
$\mb{U_r}=\mb{U_s}-\mb{U_p}$ and the drag coefficient $C_D$ is a non-linear
function of the particle-based Reynolds number, $Re_p=d_p |\mb{U_r}|/\nu_f$,
which means that $C_D$ is a complicated function of the particle diameter
$d_p$, Clift \textit{et al.} \cite{Cli_78}. For example, a very often
retained empirical form for the drag coefficient is
\begin{equation} \label{expression C_D}
C_D=
\begin{cases}
\displaystyle \frac{24}{Re_p}\left[\,1 + 0.15 Re_p^{0.687}\, \right]
 & \text{if} \; Re_p \leq 1000,  \\
0.44 & \text{if} \; Re_p \geq 1000.
\end{cases}
\end{equation}

In many papers the Saffman lift force has been considered although, strictly speaking,
this lift force is only valid in an infinite domain and, therefore, should not be considered
in the vicinity of a wall. With respect to the issue of lift forces, the situation remains
rather complex since quite a variety of different expressions have been put forward,
each time for different particle and flow descriptions, and it is difficult to gather
which ones are relevant or even whether they correspond to different lift forces or to 
different expressions of the same lift force. Yet, recently an ``optimal'' lift force, 
based on rigorous studies~\cite{McL_89,McL_91,McL_93,McL_94}, has been proposed and
seems to have helped to clarify the situation. This expression has been used in a careful 
numerical LES simulation~\cite{Wan_97} to test its importance for particle deposition
and numerical outcomes have showed only a slight reduction in the deposition rate
and mainly in the  range of small diameters. For these reasons, the lift force has not 
been included in the present study.

In some approaches, other forces are also included, namely thermophoretic and electrostatic 
forces~\cite{Kro_00,Zah_07}. Nevertheless, thermophoretic forces are important only for 
ultrafine particles in presence of a temperature gradient~\cite{Zah_07} and thus are neglected 
in the present paper, since the fluid temperature is considered uniform. Furthermore, 
electrostatic forces have a range of action so small that they can be important only for 
particles with a diameter smaller than one micron~\cite{Chen_08} and, thus, they are not
considered in the present paper, since only particles with a larger size are analysed.
Indeed, it may be quite possible to include in the particle equation of motion a rather
complete chemical force between particles and the wall, given for example by the classical 
DLVO theory that includes Van der Waals forces as well as electrostatic attractive or 
repulsive forces~\cite{Goo_04}. 
This force is important mainly in a very thin layer close to the wall 
for very small, or colloidal, particles. This expression has not been retained also because, 
in the present approach, we have chosen to concentrate mainly on the hydrodynamical effects 
on particle deposition. Thus, a simplified chemical force is actually used : there is no
chemical force inside the flow domain but when a particle hits the wall it is regarded as
being deposited, that is an infinite adhesion force is assumed.

In two-phase flow modeling, various approaches can be followed.
In this paper, we have chosen an hybrid Eulerian/Lagrangian PDF one.
We describe the continuous phase with a classical Eulerian momentum approach, 
that is the fluid phase is represented by Reynolds average
 Navier-Stokes (RANS) equations.
On the other hand, the particle phase is solved with a PDF approach
where we substitute the instantaneous exact equations
with a set of modeled instantaneous equations.
From a mathematical point of view,
these modeled equations are Langevin equations, 
that is a set of stochastic differential equations (SDEs).
A complete and rigorous presentation of this approach can be found 
elsewhere \cite{Min_04,Min_01} while 
for a general presentation of the argument of PDF modeling in turbulence
we refer to a classical reference \cite{Pop_94} 
and to the recent book of Pope \cite{Pop_00}.
The Langevin model discussed in this paper was recently proposed \cite{Min_04}
and has the form
\begin{eqnarray}
\label{eq:dxp}
dx_{p,i} &=& U_{p,i} dt \\ 
\label{eq:dUp}
dU_{p,i} &=& \frac{1}{\tau_p}(U_{s,i} - U_{p,i}) dt \\
\label{eq:dUs}
dU_{s,i} &=& -\frac{1}{\rho_f}\frac{\partial \lra{P} }{\partial x_i}\, dt
+ \left( \lra{U_{p,j}} - \lra{U_{f,j}} \right)
\frac{\partial \lra{U_{f,i}}}{\partial x_j}\, dt \nonumber \\
 &&-\frac{1}{T_{L,i}^*}
\left( U_{s,i}-\lra{U_{f,i}} \right)\, dt \nonumber \\
&& + \sqrt{ \lra{\epsilon}\left( C_0b_i \tilde{k}/k
+ \frac{2}{3}( b_i \tilde{k}/k -1) \right) }\, dW_i.
\end{eqnarray}

The crossing-trajectory effect (CTE), that is the effect due to the presence 
of external forces,
 has been modeled with the introduction of modified time-scales
 according to Csanady's analysis.
Assuming for the sake of simplicity
that the mean drift is aligned with the first coordinate axis, the
modeled expressions for the timescales are, in the longitudinal
direction:
\begin{equation}
\label{eq:CsanadyL}
T_{L,1}^{*}= \frac{T_L}
{\sqrt{ 1 + \beta^2 \displaystyle \frac{\lra{{\bf U}_r}^2}{2k/3}} }
\end{equation}
and in the transverse directions (axis labeled 2 and 3)
\begin{equation}
\label{eq:CsanadyT}
T_{L,2}^{*}= T_{L,3}^{*} = \frac{T_L}
{\sqrt{ 1 + 4\beta^2 \displaystyle \frac{ \lra{{\bf U}_r}^2}{2k/3}}}
\end{equation}
where $T_L$ represents  the Lagrangian  time-scale 
of velocity correlations and it is defined by
\begin{equation} \label{model-TL}
T_L = \frac{1}{(1/2 + 3/4 C_0)}\frac{k}{\lra{\epsilon}}~,
\end{equation}
in which $\beta$ is the ratio of the Lagrangian and the Eulerian timescales
of the fluid $\beta=T_L/T_E$, that is considered as a constant.
In the diffusion matrix we have introduced a new kinetic energy:
\begin{equation}
b_i= \frac{T_L}{T_{L,i}^*}~;~~~~ \tilde{k}= \frac{3}{2} \frac{\sum^3_{i=1}b_i\lra{u_{f,i}^2}}{\sum^3_{i=1}b_i}.
\end{equation}
All these expressions are to be regarded as being local in space and evaluated at the
particle position, that is for example $T_L=T_L (\mathbf{x}_p)$, which shows that, in
nonhomogeneous situations, the stochastic equations are non-linear. The reasoning leading
to the construction of this Langevin model as well as a discussion of the case of 
general axis direction are developped in another work~\cite{Min_01}.

It is important to underline that the solution of this set of stochastic equations
represents a Monte Carlo simulation of the underlying pdf. Therefore, this approach 
is equivalent to solving directly the corresponding equation for the pdf in the 
state-variable space. Indeed, the complete Langevin equation model for the state 
vector ${\bf Z}=({\bf x}_p,{\bf U}_p,{\bf U}_s)$ can be written 
\begin{subequations} \label{eq:sde}
\begin{align}
dx_{p,i} & =  U_{p,i}\, dt  \\
dU_{p,i} & =  A_{p,i}(t,{\bf Z})\, dt   \\
dU_{s,i} & =  A_{s,i}(t,{\bf Z})\,dt + \,  B_{s,ij}(t,{\bf Z})\;dW_j.
\end{align}
\end{subequations}
This formulation is equivalent to a Fokker-Planck equation given in
closed form for the corresponding pdf $p^L_p(t;{\bf y}_p,{\bf V}_p,{\bf V}_s)$
which is, in sample space
\begin{equation}
\label{eq:t_p}
\frac{\partial p^L_p}{\partial t} 
+  V_{p,i} \frac{\partial p^L_p}{\partial y_{p,i}} = 
- \frac{\partial}{\partial V_{p,i}}(A_{p,i}\, p^L_p\,) 
 - \frac{\partial}{\partial V_{s,i}}(A_{s,i}p^L_p\,)
+ \frac{1}{2}\frac{\partial^2}{\partial V_{s,i}\partial V_{s,j}}
           ([{\bf B}_s {\bf B}_s^T]_{ij}\, p^L_p\,)~.
\end{equation}
It can then be shown that the Eulerian MDF (mass density function) 
$F_{p}^{E}(t,\mathbf{x};\mathbf{V}_p,\mathbf{V}_s)$ satisfies the same equation 
from which the resulting (Eulerian) mean field equations can be computed~\cite{Min_01}.

Some specific characteristics of the present Langevin type of model are worth emphasizing,
particularly with respect to the simulation of small-inertia particles using an hybrid
formulation. Indeed, for very small particles (for which the mean relative drift can be
seen as negligible $ \lra{{\bf U}_r} \simeq 0$) corresponding to the limit of vanishing
inertia, $\tau_p \to 0$, also called the particle-tracer limit), the model reverts to a
Langevin model for a fluid particle since $\mathbf{U}_p \to \mathbf{U}_f$ and has the form :
\begin{subequations}
\begin{align}
dx_{f,i} &= U_{f,i} \,dt\label{eq:modelp} \\
dU_{f,i} &= -\frac{1}{\rho}\frac{\partial \lra{ P }}
{\partial x_i}\, dt 
-\frac{1}{T_L}(U_i- \lra{ U_i })dt
 + \sqrt{C_0\lra{ \epsilon }}dW_i. 
\label{eq:modelv} \end{align}
\end{subequations}
This model corresponds to the Simplified Langevin Model (SLM)~\cite{Pop_94}. 

A first important issue to consider is to be sure that the model is free of spurious drifts.
In models such as SLM, which are written as stochastic differential equations for the 
instantaneous fluid velocity $\mathbf{U}_{f}$, spurious drifts (which are related to 
spurious accumulations of fluid particles in regions of low turbulent kinetic energy)
are naturally avoided with the proper introduction of the mean-pressure 
gradient~\cite{Pop_87,Min_01}. To underline that point, it may useful to rewrite the
same model for the fluid particle velocity fluctuating component 
$\mathbf{u}_f=\mathbf{U}_f - \lra{\mathbf{U}_f}$ which is
\begin{subequations}
\begin{align}
dx_{f,i} &= ( \lra{U_{f,i}} + u_{f,i}) \,dt \\
du_{f,i} &= \frac{ \partial \lra{ u_{f,i}u_{f,k} } }{\partial x_k}\, dt 
 -u_{f,k}\frac{ \partial \lra{U_{f,i}} }{\partial x_k}\, dt 
-\frac{u_{f,i}}{T_L}\, dt + \sqrt{C_0\lra{ \epsilon }}dW_i. 
\end{align}
\end{subequations}
Thus, in non-homogeneous situations, the increments of the fluctuating velocity components
along a Lagrangian trajectory have a non-zero value, due to the first term on the rhs of
the last equation (there is an underlying difference between means taken along fluid 
particle trajectory, in a Lagrangian setting, and mean values at a fixed point, in an
Eulerian setting, which for the fluctuating velocity is of course zero). Although surprising
at first sight, this term is absolutely necessary so as to be able to respect the 
incompressibility constraint which states that a uniform fluid particle concentration
should remain uniform even in a non-homogeneous situation~\cite{McI_92,Pop_87,Min_01}. 
However, models (for example some models of the random-walk type) that simply add to 
the mean fluid velocity a fluctuating component that has a zero-mean value (thus 
confusing Lagrangian and Eulerian averaging operators) are equivalent to models where 
an artificial drift velocity is implicitely added in the correct equation, namely 
$v_{d,i}= - \partial \lra{ u_{f,i}u_{f,k}}/\partial x_k$. In the channel flow approximation, 
where $v_d= - d\lra{ v^2 }/d y$ in the direction normal to the wall,
this amounts to adding  a spurious drift that artifically drives fluid particle
away from the wall, thereby reducing the possibility of small-particle deposition.

A second relevant issue is the consistency of Eulerian and Lagrangian turbulence modeling.
Indeed, in terms of Eulerian mean equations, the SLM model is equivalent to the following 
set of equations~\cite{Pop_94b} :
\begin{eqnarray}
\label{eq:continuity}
&&\frac{\partial \lra{ U_i }}{\partial x_i} =0  \\
&&\frac{\partial \lra{ U_i }}{\partial t} + 
\lra{ U_j} \frac{\partial \lra{ U_i }}{\partial x_j}
+ \frac{\partial \lra{ u_iu_j }}{\partial x_j} =
-\frac{1}{\rho}\frac{\partial \lra{ P }}{\partial x_i}  \\
&&\frac{\partial \lra{ u_iu_j }}{\partial t}
+ \lra{ U_k }
\frac{\partial \lra{ u_iu_j }}{\partial x_k}
+\frac{\partial \lra{ u_iu_ju_k }}{\partial x_k} = 
-\lra{ u_iu_k }\frac{\partial \lra{ U_j }}
{\partial x_k} - \lra{ u_ju_k }\frac{\partial \lra{ U_i}}
{\partial x_k} \nonumber \\
&& \hspace{7cm} -\frac{2}{T_L}\lra{ u_iu_j } + C_0\lra{ \epsilon } \delta_{ij}.
\label{eq:rans}
\end{eqnarray}
Using the expression retained for $T_L$ in Eq.~(\ref{model-TL}), the transport equation for
the second-order moments can be re-expressed as :
\begin{eqnarray}
&&\frac{\partial \lra{ u_iu_j }}{\partial t} + \lra{ U_k }
\frac{\partial \lra{ u_iu_j }}{\partial x_k}
+\frac{\partial \lra{ u_iu_ju_k }}{\partial x_k} = 
-\lra{ u_iu_k }\frac{\partial \lra{ U_j }}
{\partial x_k} - \lra{ u_ju_k }\frac{\partial \lra{ U_i}}
{\partial x_k} \nonumber \\
&& \hspace{7cm} -(1 + \frac{3}{2}C_0)\left( \lra{u_iu_j} - \frac{2}{3}k \delta_{ij} \right)
 - \frac{2}{3}\delta_{ij}\lra{ \epsilon }.
\end{eqnarray}
This shows that the SLM corresponds to a $R_{ij}-\epsilon$ Rotta model~\cite{Pop_94b}. 
It is important to underline that the complete stochastic model, which is based on an
assumption of an isotropic return-to-equilibrium term for the closure of the 
pressure-strain correlation, is not isotropic even in the asymptotic case of tracer 
particles, that is for the fluid case. Yet, as it transpires from its name, the SLM is 
perhaps the simplest possible stochastic model consistent with classical Reynolds-stress 
second-order modeling and its capacity to reproduce high anisotropy, such as in the 
near-wall turbulent boundary layer, is limited~\cite{Min_99}. It is possible to replace
the simple return-to-equilibrium term in Eq.~(\ref{eq:modelv}) by a more general matrix $G_{ij}$
which is a function of local fluid mean velocity gradients~\cite{Pop_94b,Min_99} so as to 
retrieve more complex Reynolds-stress models for $\lra{ u_iu_j }$ which may improve 
numerical predictions in highly-anisotropic regions. New complete (and more complex) Langevin 
models have also been recently put forward with down-to-the-wall integration and are able 
to reproduce the high-anisotropy of the Reynolds-stress quite well~\cite{War_04}. 
However, in the present context, we are using an hybrid formulation and we believe that,
before resorting to more involved models, it is important to stress the consistency issue. 
Indeed, in such a formulation, one turbulence model is used in the Eulerian part for the 
prediction of the fluid mean fields such as the mean velocity and Reynolds-stress. 
These fluid mean fields are provided to the Lagrangian solver in 
Eqs.~(\ref{eq:dxp})-(\ref{eq:dUs}) which also corresponds to a turbulence model, as it 
was just underlined. For small-inertia particle,
we have therefore a duplicate turbulence model and it is very important to ensure that
these two turbulence models be as consistent as possible~\cite{Pop_00,Chi_06}. 
Indeed, it has been shown that to couple models which correspond to different turbulence models 
(for instance DNS and the present Lagrangian model) may introduce some inconsistencies at the 
level of particle equations and, thus, may lead to unphysical results in particular for
the numerical prediction of wall-normal stress, say $\lra{v^2}$, which is important if we are
to simulate particle fluxes towards the walls~\cite{Chi_06,Par_08}.
Therefore, as a first step, we have retained a simple version, namely the SLM model, which
is consistent with usual Reynolds-stress models as a kind of sound basis for the numerical
investigations on particle deposition though it is clear that, at least for the prediction
of fluid mean quantities, this leaves room for improvement by using more complex Langevin ideas. 

\section{Numerical results}
\label{sec:results}

In this section, we present numerical results for the deposition of particles
in a vertical pipe flow at a Reynolds number of 10 000, which corresponds to
the experiment of Liu and Agarwal (1974)~\cite{Liu_74}.

In order to describe the particle phase, 10,000 individual particles 
($920 kg/m^3$ in density)
of 10 diameters ($1.4-68.5 \mu m$) are released in the gas flow.
In table \ref{tab:taup_dam}, we report the relation between particle diameters
and characteristic response times, 
based on the definition given in Eq.~(\ref{taup}).
The numerical integration of the Langevin equations describing the particle phase is 
fully described in a recent paper~\cite{Min_03}.
To compute the deposition velocity,
we evaluate $F$, the fraction of particles remaining in the flow,
as a function of the axial position $x$ \cite{Mat_00}. 
$F$ is calculated by counting the number of particles 
that reach the sampling cross-section and it is defined by
$$F =  \frac{number~ of~ crossing~ particles}{total~ number~ of~ released~ particles}$$
The particle deposition velocity is then computed as follows \cite{Mat_00}
\begin{equation}
k_p = \frac{U_f d_t}{4(x_2 - x_1)}ln\frac{F_1}{F_2}~,
\end{equation}
where $d_t$ is the diameter of the pipe and $F_i$ is particle fraction value
at the $i-th$ sampling section.
As previously explained, pure-deposition boundary conditions are applied for the particles,
that is particles touching the wall are considered as being deposited and are removed from 
the domain. For the test-case simulated in this work, the aerosol flow is considered as 
dilute and, thus, interactions between turbulence and particles are only one-way.

\subsection{Mean fluid value predictions}
\label{sec:eul} 

\begin{figure}[htbp]
\begin{center}
\epsfig{file=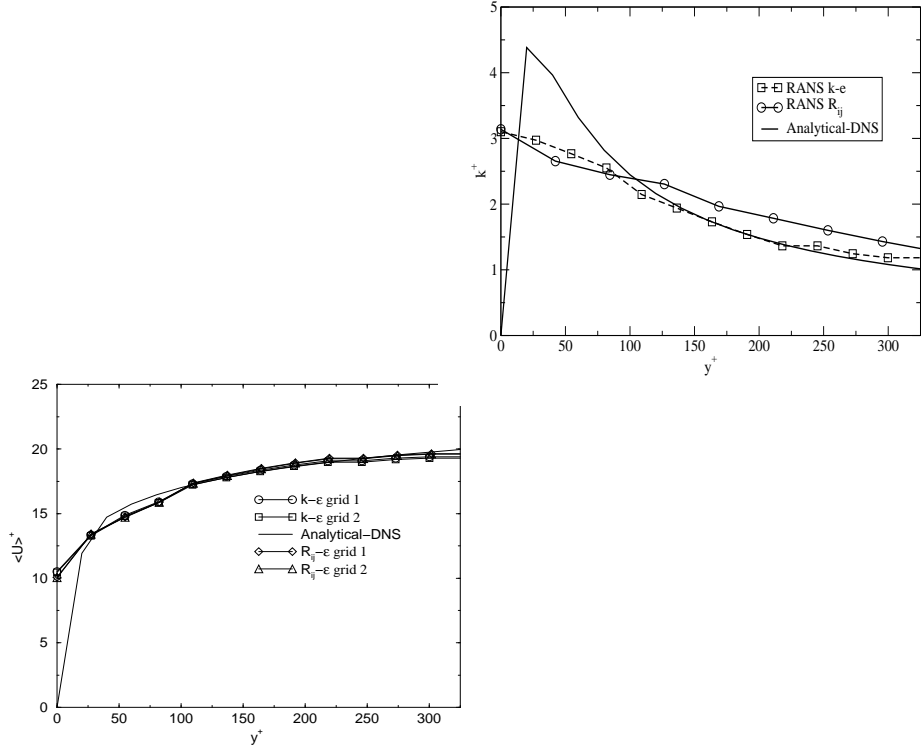,width=5.truecm,height=6cm, angle=-90}
\epsfig{file=eul2.eps,height=5.truecm,width=6cm}\vspace{1.cm}
\caption{(a): The mean velocity of the fluid phase is shown versus the distance from the wall. All quantities are made non-dimensional with the wall friction velocity and viscosity. 5 curves are shown: the ones labeled by ``grid 1'' represent the results obtained with the grid used trhroughout this paper (168000 cells) with both turbulence models used, namely $k-\epsilon$ and $R_{ij}-\epsilon$. The curves labelled ``grid 2''  are obtained on a grid with a resolution doubled in radial and azimuthal direction with both turbulence models. Results are very similar and therefore the grid-independence can be considered reached. The last curve shows the analytical curve that fits DNS results. 
(b): Adimensional turbulent kinetic energy is shown versus the adimensional distance from the wall.  Analytical/DNS results are also shown for comparison. As expected, the peak of the kinetic energy is under-estimated by present turbulent models. Globally speaking, our numerical results are in line with standard performances of these models \cite{Pop_00}.}
\label{fig:eul}
\end{center}
\end{figure}

Although the purpose of this work is to analyse Lagrangian modeling for particle deposition, 
we first show some Eulerian results for the sake of completeness. Indeed, in the hybrid 
approach, the first step to be carried out is to evaluate the mean fluid variables 
which are included in the Lagrangian model, see eqs.~(\ref{eq:dxp})-(\ref{eq:dUs}).
The pipe test-case considered in this work has been solved on an unstructured 
grid composed by $168000$ points, that is $12\times28\times500$ points in the three directions.
For all computations, we have used the Reynolds Averaged Navier Stokes (RANS) free code ``Saturne'',
which is an in-house code developed at Electricit\'e de France. All details about this 
rather classical computational fluid dynamics code can be found elsewhere~\cite{Saturne}.
Grid-independence has been assured, as shown in fig.~\ref{fig:eul}a.
Wall-boundary conditions have been imposed through classical wall-functions, with the 
first grid-point put at $y^+\approx50$~\cite{Wilcox}.
At the inlet, the mean velocity is imposed uniform and equal to the bulk velocity $U_m$ 
given by Reynolds number $Re=\frac{U_mH}{\nu}$, where $H$ is the Pipe diameter.
No variation with radial position is present. 
Steady state was obtained after some douwnstream distance; nevertheless, in order to be sure 
that inlet conditions have no effect on our results, only second half of the channel has been considered. Outflow conditions were imposed at the outlet. 

Standard $k-\epsilon$  and $R_{ij}$ turbulence models have been used. Results are in line 
with the known performance of these models in wall flows and very similar results have been 
obtained with both models.  In fig.\ref{fig:eul}a, the mean velocity in the axial direction 
obtained with both models for two grids are shown and compared against analytical results~\cite{Mon_75}, 
which fit very well DNS results of comparable Reynolds number~\cite{Man_88}.
In fig.\ref{fig:eul}b, the turbulent energy obtained with both models is shown.

\subsection{Results with the standard model}
\label{sec:test} 

\begin{figure}[htbp]
\begin{center}
\epsfig{file=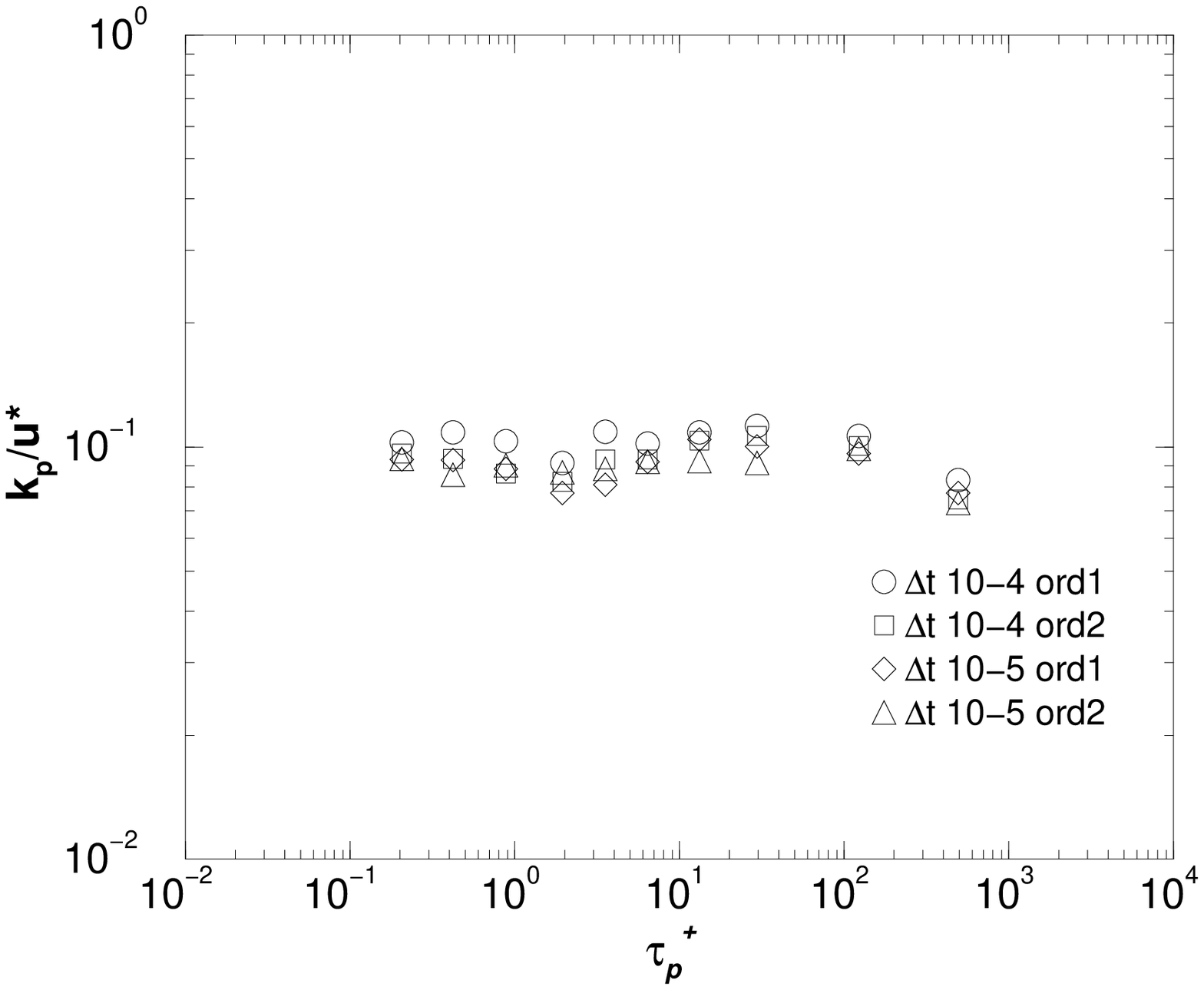,height=7.truecm}
\epsfig{file=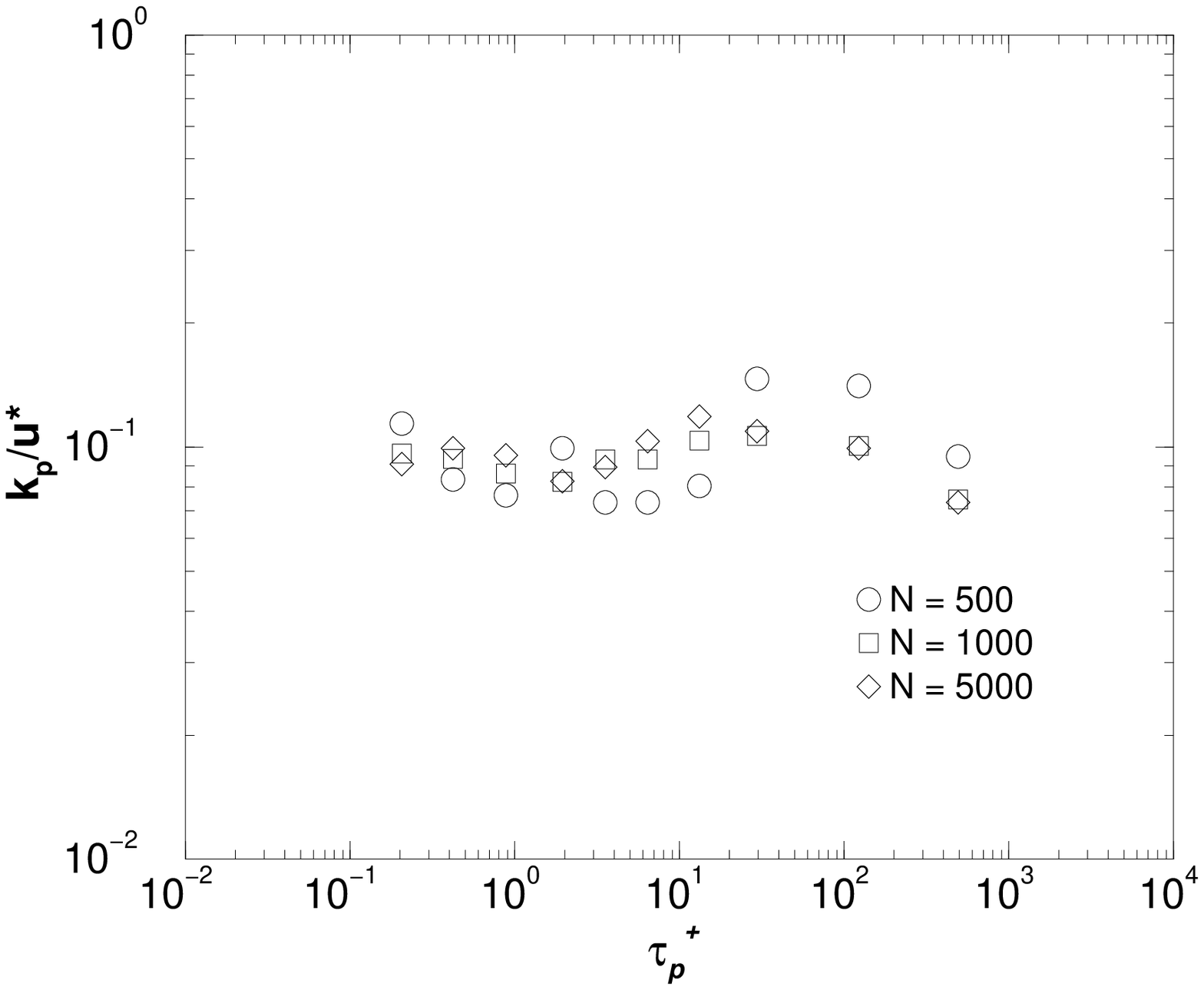,height=7.truecm}
\caption{(a): Analysis of the convergence of the numerical scheme with the order and the time-step. The deposition velocity (y axis) is shown versus adimensional particle response time for different numerical configurations. For the test-case studied in this work, tt is seen that the for $\Delta t=10^{-4}$ the results have reached the convergence and that the order of the scheme is not a key.
(b): Analysis of the particle number influence on the results of deposition velocity. The Deposition velocity (y axis) is shown versus adimensional particle response time for some configurations with a different number of particles. It is seen that $N=500$ is enough to guarantee the independence of the results from the number of particles used.}
\label{dt}
\end{center}
\end{figure}

First of all, numerical tests have been performed to check that the
results were independent of the values of numerical parameters, in particular
the number of particles and the time-step.
Results are shown in
Fig.~\ref{dt}, where a standard $k - \epsilon$ turbulence model
was applied for the fluid in the Eulerian solver. It is seen that numerical
independence with respect of the time step is reached for the two numerical
schemes (of order 1 and order 2), and that the influence of the order of
convergence of the schemes is negligible. Based on these results, a time-step
of $\Delta t = 10^{-4}s$ was chosen for the different calculations, while
keeping the second-order algorithm.
The independence of the deposition velocity 
with respect to the time-step  illustrates that the numerical scheme 
is stable for the whole range of particles.
It is well knowné \cite{Pei_06} that, for a given time step,
the stochastic equations for turbulent particle become stiff
for  small diameters and near the wall.
If this mathematical characteristic is not well addressed with an appropriate
numerical scheme,  the stiffness problem imposes the use of very small time-step 
in order to prevent the presence of numerical instabilitiesé \cite{Mat_00,Kro_00}
which  may also lead to the use of an unphysical time-step.
Thus, the present algorithm appears as satisfactory for particle deposition
computations.

Apart from numerical errors due to the time-accuracy of the numerical
scheme, an analysis of the statistical error has been carried out.
Since particle deposition velocities are calculated by a Monte Carlo method,
it is important to check that the number of particles (which represents
samples of the pdf) is sufficiently high so that statistical error
is limited. In Fig.~\ref{dt}, we present also results obtained 
with three different values of N, which is the number of particles
used for each class of diameter : $N = 500, 1000$ and $5000$.
As it appears, although results change very slightly with increasing $N$,
there is no clear difference between these results and
it seems that 500 particles for each class of diameter is already high enough.
However, we have chosen for further simulations the value of
$N = 1000$ particles for each class of diameter,
in order to reduce statistical noise.

\begin{figure}[htbp]
\begin{center}
\epsfig{file=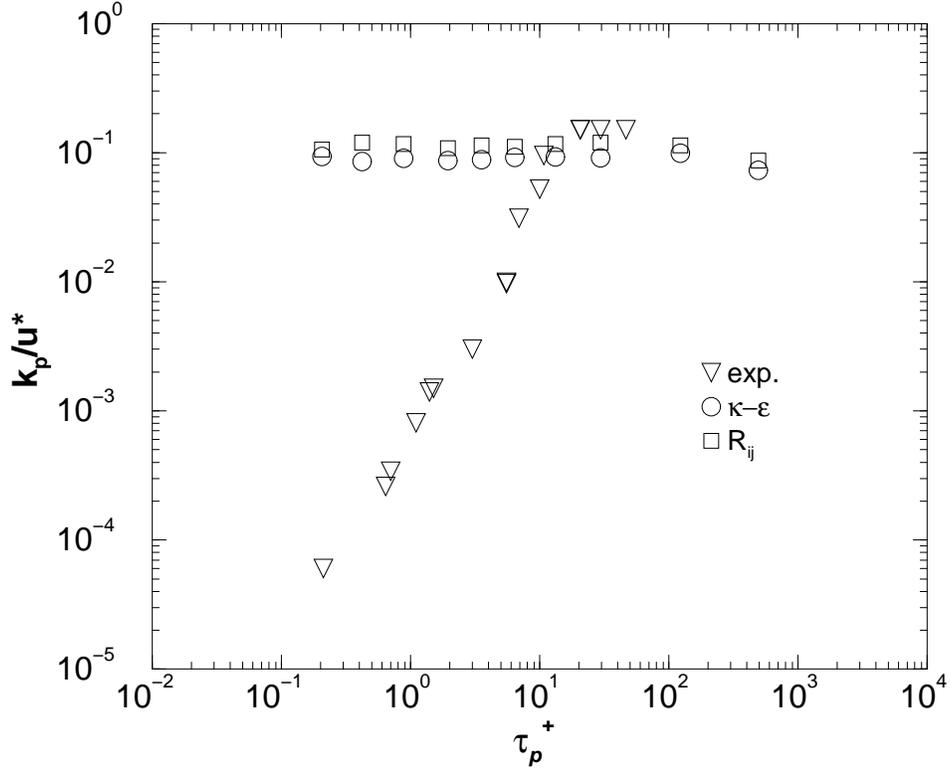,height=11.5truecm}
\caption{Analysis of the influence of the turbulence model used for the fluid phase. The deposition velocity obtained with different turbulent models is shown: $k-\epsilon$ circles,  $R_{ij}-\epsilon$ squares. The results are almost indistinguishable, showing that the stochastic model used in this work for the description of the particle phase is not much affected by the turbulence model used for the computation of the mean fluid variables.}
\label{turb_dep}
\end{center}
\end{figure}

In figure \ref{turb_dep}, results obtained with the standard PDF model,
Eqs.~(\ref{eq:dxp})-(\ref{eq:dUs}),
can be compared with experimental data.
Two different turbulent models for the simulation of the continuous phase were used, 
namely the standard $k - \epsilon$ and $R_{ij}-\epsilon$ models,
both with wall-function boundary conditions.
The difference between the simulations performed with the two 
different turbulence models is negligible.
This is not too surprising, since in this test-case the two models give 
similar mean fluid profiles, as shown in previous section.
These results are coherent with those obtained in an analogous configuration
by Schuen \cite{Sch_87}.

It is possible to divide the results in two main categories.
For heavy particles ($\tau_p^+ > 10$), the model prediction agrees very well
with experiments.
On the contrary, for light particles ($\tau_p^+ < 10$), the deposition velocities
remain at the same level as for heavy particles,
and are therefore strongly overestimated.
This fact shows that the stochastic model
proposed, in its standard form, 
is not suitable to simulate  deposition phenomena in this range.
In particular, there is no appreciable difference
between the two categories ($\tau_p^+ < 10$ and $\tau_p^+ > 10$)
in the standard form of the model,
 while in reality 
deposition velocity diminishes by three order of magnitude.

This seems to be in line with experimental and DNS results, 
heavy particles are not much affected by near-wall boundary layer 
and by the specific features of the instantaneous turbulent structures
and, thus, the general model provides an adequate description.
On the contrary, for light particles, the physical mechanism of deposition
changes, with a growing importance of turbulent structures and 
near-wall physics in general.
The standard form of the PDF model is not sensitive to this change,
giving for particles in the whole spectrum of diameters almost 
the same results.

\subsection{Influence of mean fluid profiles}
\label{sec:mean} 

As previously mentioned, in the standard form of the PDF model, 
the fluid boundary layer is simulated with the wall-function approach.
This method assures a reasonable approximate mean fluid profile
in the logarithmic region without solving explicitly the viscous sub-layer.
Nevertheless, given that the results obtained with the standard model
are not satisfying,
a question arises : is the prediction of particle
deposition velocity sensitive to changes in the fluid mean fields ?
Or, to be more precise, can we improve predictions
by changing the value of mean-quality profiles and of parameters 
(such as $T_L$) that enter in the Langevin equation,
while keeping the same form of the model?

\begin{figure}[!h]
\begin{center}
\epsfig{file=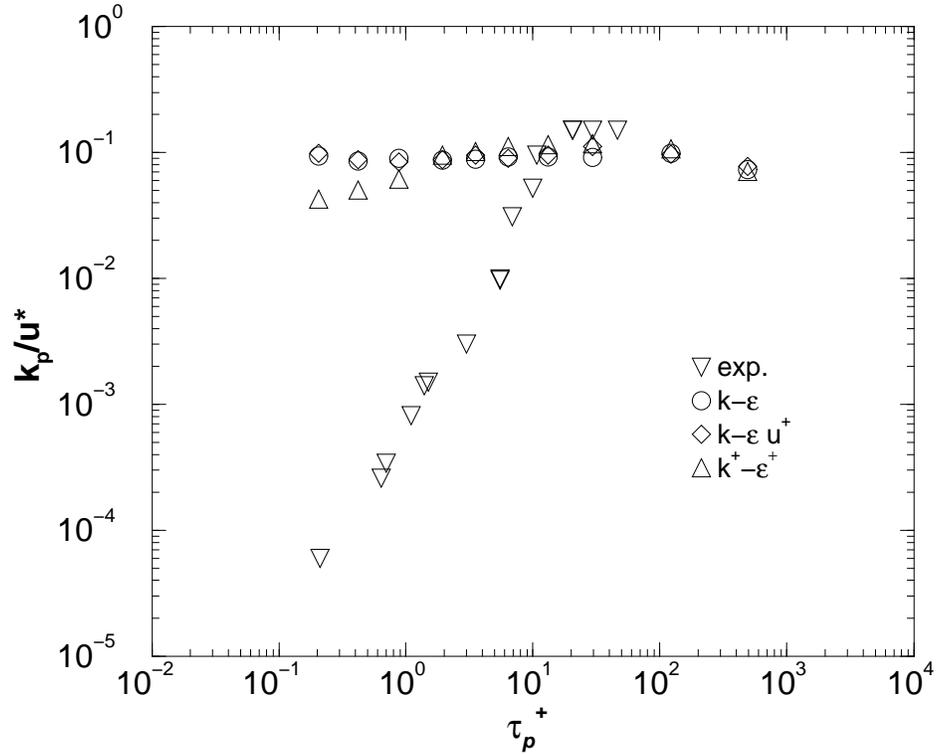,height=10.truecm}
\caption{Deposition velocity with different fluid velocity profiles.
Triangles down are the experimental value. 
Circles are obtained with all mean fluid quantities given by
$k-\epsilon$ model (as in the previous picture).
Diamond curve is obtained imposing  axial mean fluid velocity 
given by the law-of-the-wall equations $\lra{U_{f,i}} = u^+$.
For the result indicated by diamonds with $k^+-\epsilon^+$,
the mean fluid velocity is always given by the same law,
and also turbulent kinetic energy ($k$) 
and turbulent dissipation rate ($\epsilon$) are curve-fitted to the DNS data
\cite{Mat_00},
thus all mean fluid profiles are exact.}
\label{viscous}
\end{center}
\end{figure}

The influence of such approximations was recently investigated 
by other Lagrangian models~\cite{Mat_00}.
Following the same reasoning, we have carried out simulations 
where the computed mean fluid fields are replaced by given ones
in the whole domain and, consequently, 
wall-function boundary conditions are suppressed.
In the chosen test-case, 
analytical solutions for the mean fluid fields ($\lra{{\bf U}},k,\lra{\epsilon}$)
can be found~\cite{Mon_75} and DNS data are also available 
for the entire region of simulation.
It is important to underline that, for this particalar numerical study,
grid resolution has been largely improved, with a grid spacing near-to-the-wall of
$\Delta y^+ \approx 1$. This should assure that mean variations are captured with sufficient detail.
The aim of this substitution is to  make a sensitivity analysis 
of the standard model with respect to  mean fluid quantities,
because one might expect that smaller particles are more 
sensitive to the rapid variations of mean quantities 
expected in near-wall layer.
In figure~\ref{viscous}, we present results for different tests.
A first sensitivity test has been carried out with imposing  the axial mean fluid velocity 
given by the law-of-the-wall equations $\lra{U_{f,i}} = u^+$,
that is we have used the theoretical analytical value for this variable.
In the second test,
again the mean fluid velocity has been computed through  this  law,
but we have also used  turbulent kinetic energy ($k$) 
and turbulent dissipation rate ($\epsilon$) curve-fitted to the DNS data~\cite{Mat_00}, 
thus all mean fluid profiles used in this test are ``exact'',
in the sense that they are either given by the analytical solution (mean velocity)
or by DNS simulations.

The influence of the fluid mean profiles on the predicted values is limited and the 
model does not appear to be sensitive to them. The deposition rate remains over-predicted 
by the model for light particles. This may be explained by the fact that the effect of 
the new mean fluid profiles is concentrated in a thin region. Most important quantities 
are expected to be the turbulent kinetic energy and the wall-normal stress~\cite{Par_08}.
Nevertheless, turbulent kinetic energy decreases only from $y^+ \approx 10$,
where it reaches its maximum in correspondence with peak production.
Wall-normal stress peaks further but yet near-to-the-wall, at about 
$y^+ \approx 50$ \cite{Kim_87}. 
The resulting effect is not easy to be foreseen 
and it may be negligible with respect to the overall effect of
migration of particles towards the wall due to the net mean flow.
In order to further support this argument, 
we have computed the mean near-wall residence time (in the layer $y^+ < 30$)
of deposited particles, for each class of diameters.
We have chosen to monitor the particle residence-time because
this quantity has been found to properly  distinguish 
different deposition mechanisms~\cite{Nar_03}.
In Table \ref{sej1}, 
the results obtained for each class of diameters are given
for the simulation with all exact fluid profiles.
For the sake of clarity, the residence time is always expressed 
in nondimensional wall- units 
(i.e. normalized using the kinematic viscosity and the friction
velocity).
In the model, all  particles, regardless of  their diameter,
are found as deposing by the free-flight mechanism, 
that is with a small near-wall residence time.
Furthermore, the residence time grows slightly with diameters.
This fact shows that particles are dominated by the migratory flux
and light particles are even faster than the biggest ones 
to reach walls, 
since the acceleration on particles is proportional to the inverse of diameter.

A first conclusion can be drawn: in the absence of a representation of turbulent 
coherent structures which can  trap particles in the near-wall region and which 
describe correctly the mechanisms of deposition, the mean fluid profiles are not
found to be a significant factor.
In some previous works~\cite{Mat_00,Par_08,Wan_99}, it was experienced that 
the introduction of exact mean fluid quantities improved the performance of discrete 
Lagrangian models. However, the same tendency has not been observed in the present work.
With respect to this point, it may be worth remembering that the attention in these works 
was mainly devoted to the analysis of the Eulerian part of the hybrid approach and that, 
very often, a standard ``random-walk'' model was used for the Lagrangian part.
In the present work, a rather complementary point of view has been followed, where the 
emphasis was put on the Lagrangian model and, more specifically, on the consistency between 
the Eulerian and Lagrangian formulations in the fluid limit. The theoretical issues
related to this consistency question have already been developed in the previous part 
but they are further compounded by similar numerical issues, so that we believe that
it is important to address carefully several aspects in practical computations while
testing the sensitivity to mean fluid profiles :
\begin{enumerate}
\item[(a)] Lagrangian models can be affected by spurious drifts~\cite{Pop_87,McI_92,Min_01}, 
as discussed in the previous section, which may correspond to an artificial force which 
pushes small particles away from wall. It must be ensured that a correct mean pressure-gradient
is correctly introduced before pursuing further tests~\cite{Chi_06}.
\item[(b)] Lagrangian models are written as stochastic differential equations (SDE) whose 
numerical integration is more subtle than classical ordinary differential equations (ODE). 
A straightforward approach based upon classical numerical schemes for ODE can lead also 
to the existence of spurious drifts, now of numerical origin~\cite{Pei_06,Min_03b}. 
\item[(c)] In Lagrangian simulations, if standard numerical schemes are used, a 
very small time-step is required near the wall to guarantee numerical stability. 
This may lead to an unphysical behaviour, since present stochastic models are based upon 
the hypothesis that the time-step is much greater than the Kolmogorov time-scale 
$\Delta t \gg \tau_{\eta}$. 
\item[(d)] It has been found that it is important to ensure that the turbulence Eulerian model 
and the Lagrangian one are as consistent as possible~\cite{Chi_06,Mur_00}. The lack of 
consistency may also lead to unphysical results, at least for the limit case of very small 
particles~\cite{Chi_06}. In particular, even with the exact mean profiles, the present 
Langevin model do not reproduce exactly the Reynolds stress, for example the wall-normal 
stress may be slightly underestimated, and thus this can limit the effect
of the introduction of better Eulerian predictions. 
\end{enumerate}

With the previous issues in mind and given the results obtained in this section, we 
propose to retain the present Langevin model, but to implement it with a simple phenomenological 
model to account for some of the near-wall physical mechanisms due to coherent structures.
The purpose of this new phenomenological model is two-fold: first to improve the model 
predictions in a ad-hoc but simple manner and, second, to investigate whether modeling more 
explicitly particle interactions with near-wall coherent structures is a direction worth pursuing.

\subsection{Phenomenological model for coherent structures}
\label{sec:phemod} 

The turbulent near-wall structures have been found to have a main role 
on the mechanism of particle deposition \cite{Mar_02,Nar_03}.
For our purpose, the most interesting aspect is that depositing particles 
can be divided into two categories.
In the first one, particles with large wall-normal  
velocity and small near-wall residence time,
deposit mainly by the {\it free-flight} mechanism. 
In the second one, for particles with negligible wall-normal velocity 
and large near-wall residence time,
the {\it diffusional} mechanism is the most important one.

More specifically, for light particles ($\tau_p^+ < 10$) the 
{\it diffusional} mechanism is shown to represent
the sole mechanism useful to deposition,
while its importance decreases as particles become heavier.
Yet, Narayanan et al. \cite{Nar_03} show
that the diffusional deposition mechanism
still remains quantitatively important for heavy particles, 
at least in the intermediate range which is considered there.
For example, for particles with $\tau_p^+=15$, only about 40 \%
of the particles are expected to deposit by the free-flight mechanism.
However, for very heavy particles ($\tau_p^+ \gg 10$)
the {\it free-flight} mechanism is expected to become the dominant one.
From a physical point of view, the different behavior 
between particles of different inertia can be explained
in terms of the interaction with coherent structures in the 
near-wall region, notably with sweeps and ejections.
Light particles remain trapped for a long time 
by these structures in a thin region ($y^+ < 3$) and deposit
only by diffusion. Heavy particles, with a high enough wall-normal velocity, 
go through this region without being influenced and deposit 
after a small residence time.

Although  diffusional deposition was found to be still important
in the small range of diameters analyzed in DNS simulations,
we can propose the following picture : heavy particles
($\tau_p^+ > 10$) deposit by the {\it free-flight} mechanism, 
while light particles ($\tau_p^+ < 10$) deposit by the 
{\it diffusional} mechanism.
Though this represents a rough approximation,
it can be considered as reasonable for the construction of a 
simple model. In effect, with our standard  PDF model,
heavy particles ($\tau_p^+ > 10$) are well treated,
see Fig.~\ref{turb_dep}. Therefore, for particles with 
($\tau_p^+ > 10$) we do not need any model modification.
For light particles the situation is completely different.
Evidently, some instantaneous features of
coherent structures are not well represented 
in our standard PDF model,
so the idea is to add the main effects by new {\it ad-hoc} terms.

The results  obtained by recent DNS computations suggest 
that the  residence time of particles in near-wall region
represents the most important parameter.
Given this, we propose a simple phenomenological model which 
covers the whole range of diameter (heavy and light particles).
The model introduces the notion of a residence time scale in the near-wall 
region for each class of diameter, say $T_s(d_p)$.
The characteristic time scale $T_s(d)$
is function of particle diameters, 
and we propose to model it with the simple form
\begin{equation}
T_s = T_0 \exp\left(- \frac{d_p}{D_0} \right)~.
\label{dep_mod2}
\end{equation}
This form is based on the dimensional guess 
$\frac{dT_s}{dd_p}=-T_s/D_0$ and it is chosen because 
it gives the good monotonic and asymptotic behavior.
The two parameters $D_0$ and $T_0$, not a priori known, can be 
extracted from the two values investigated by DNS 
($\tau_p^+ = 5$ and $\tau_p^+ = 15$).
Although DNS gives only the statistical distribution
of this quantity and not a single value,
it is possible to roughly deduce from DNS results
a mean value of $T^+_s \sim 10^3$, for $\tau_p^+ = 5$,
and $T^+_s \sim 10^2$, for $\tau_p^+ = 15$.
On the basis of these results,
the parameter are evaluated to be 
$T^+_0 = 700~,~D^+_0 = 2.3 \times 10^{-4}$ in adimensional
wall units.
Though these results have been deduced from a single DNS computation
at a given Reynolds-number, they are computed from adimensional quantities related 
to wall ones, which are known to have almost universal character \cite{Pop_00},
and the present estimates are assumed  to have some general validity.

Since our model is aimed at introducing features of coherent structures 
whose  influence is limited to a thin region
near the wall \cite{Nar_03}, 
it seems reasonable to apply it in the numerical simulations
by imposing {\it ad-hoc} boundary conditions :
when a particle hits a wall, it deposits only if 
its residence time in the near-wall region (defined as the zone $y^+<30$)
is greater than $T_s$.
Otherwise, it remains at the wall and its  velocity is put to
zero, but it can be reentrained and move again in the flow.
These boundary conditions are applied to each class of particle diameters.

To sum up, the complete Langevin PDF model proposed is as follows
\begin{equation}
\left .
\begin{split}
dx_{p,i} & = U_{p,i}\, dt \\
dU_{p,i} & = \frac{1}{\tau_p}\,(U_{s,i} - U_{p,i})\, dt  \\
dU_{s,i} & = A_{s,i}(t,{\bf Z})\,dt + \,  B_{s,ij}(t,{\bf Z})\;dW_j.
\end{split} \right\} \text{SDE model}
\end{equation}
\begin{equation}
\left .
\begin{split}
\text{Particle deposition}~; &\; \text{if}~ T_p > T_s \\
U_{p,i}=0~, \; x_{p,i} = 0~;   &\; \text{if}~ T_p < T_s 
\end{split} \right\} \text{Particle B.C.}
\end{equation}
where $T_p$ represents the  residence time of the given particle 
in the near-wall layer $y^+<30$.

In our picture, heavy particles ($\tau_p^+ > 10$) 
deposit each time they reach the wall,
since the residence time scale tends to zero rapidly with 
particle diameter.
On the other hand, light particles ($\tau_p^+ < 10$) deposit only if 
they remain in the near-wall region for a sufficient time. 

\begin{figure}[!h]
\begin{center}
\epsfig{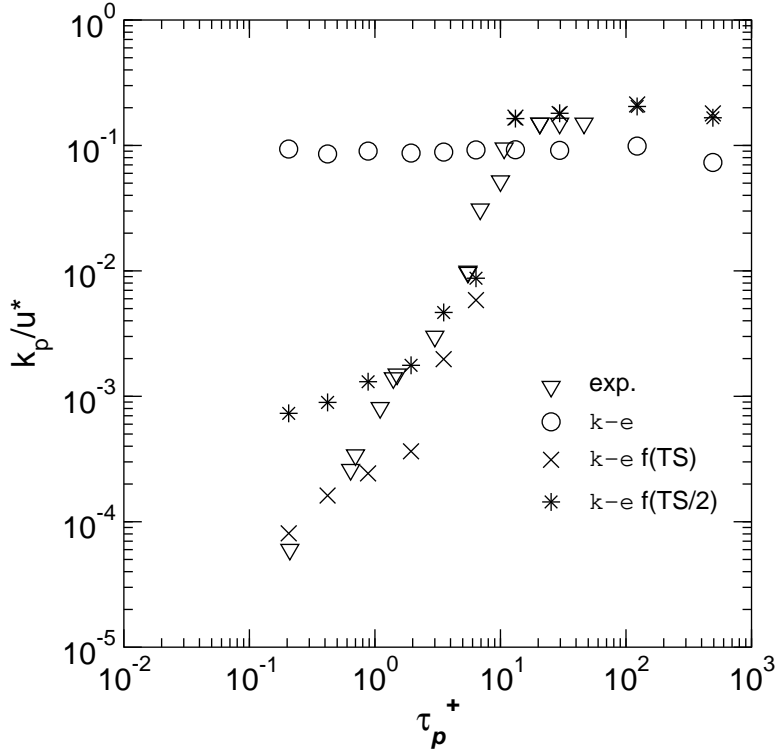}
\vspace{0.5cm}
\caption{Deposition rate velocity for the different model used. In all numerical  cases the continous phase is solved via standard $k-\epsilon$ model. Experimental results are given for refernce (triangle down). The standard results are indicated by the curve labeled with $k-e$ (circles). The results obtained with the new phenomenological model are shown by the curve indicated by k-e f(TS) (crosses). In this case, particles can deposit only if they have remained in a small near-wall layer for a time given by the function f(TS), which is derived from DNS data. The last curve (stars) indicates the results obtained with the new phenomenological model but letting particles deposit even if they have remained only the half $T_s/2$ of the residence-time given by the function f(TS) derived from the DNS data. The results obtained with the phenomenological model with the residence-time computed through the function derived from DNS are in good agreement with experimental results, in particular small particles deposit only rarely.}
\label{bc}
\end{center}
\end{figure}

In Fig. \ref{bc}, the results obtained with the new model
are represented by the curve indicated by $f(TS)$. A good agreement with
experimental data is retrieved, and in particular the sharp decrease
of the deposition velocity for light particles is correctly reproduced.
In the same figure, we present also a second curve indicated by $f(TS/2)$,
which represents the results obtained by using in the model
a residence time scale equal to $T_s/2$.
These results indicate that the dependence on the residence time 
is critical for lighter particles.
In fact, in this particular test-case considered 
it represents the main effect.

\begin{figure}[!h]
\begin{center}
\epsfig{file=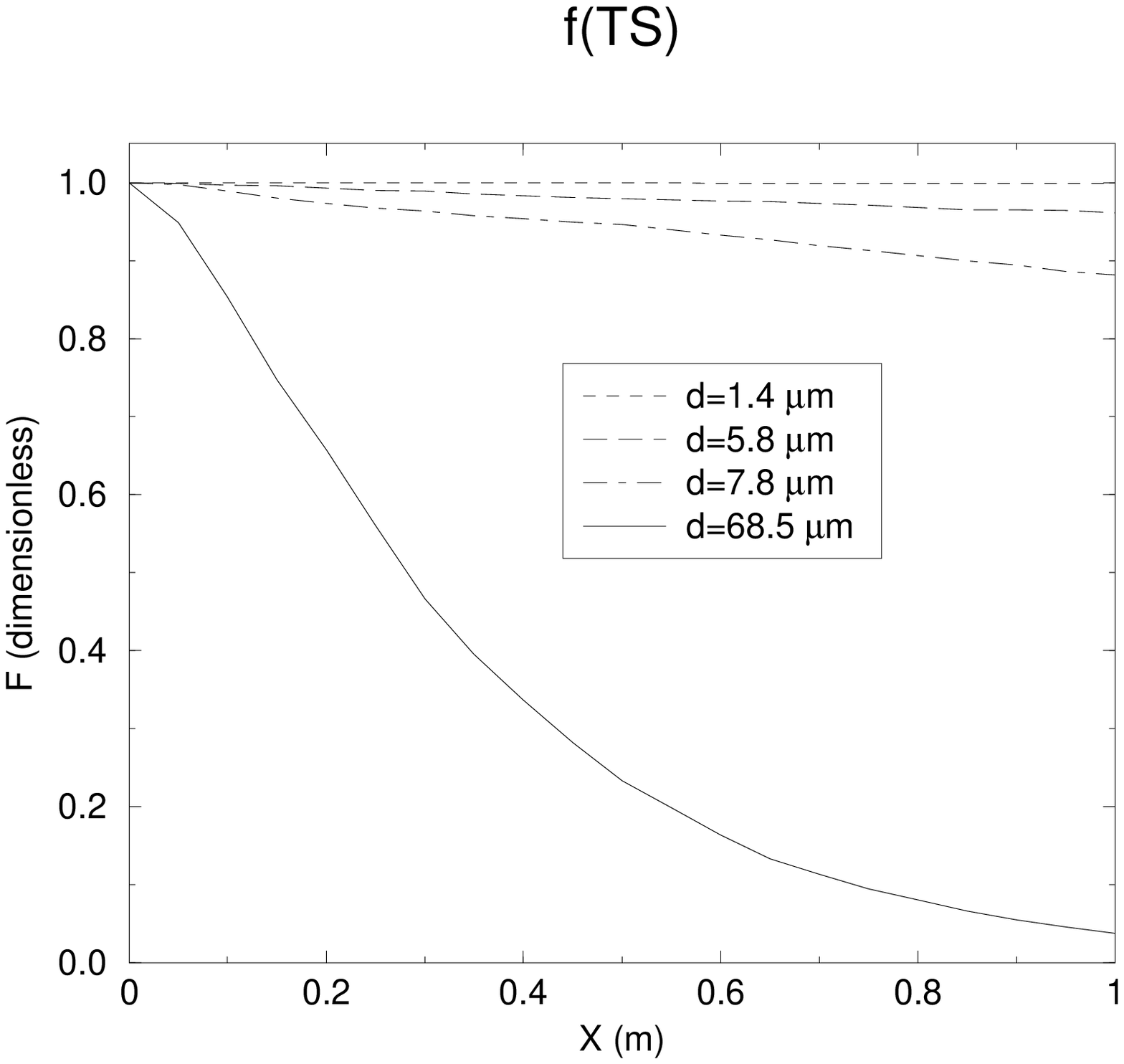,height=7.truecm,width=7.5truecm}
\epsfig{file=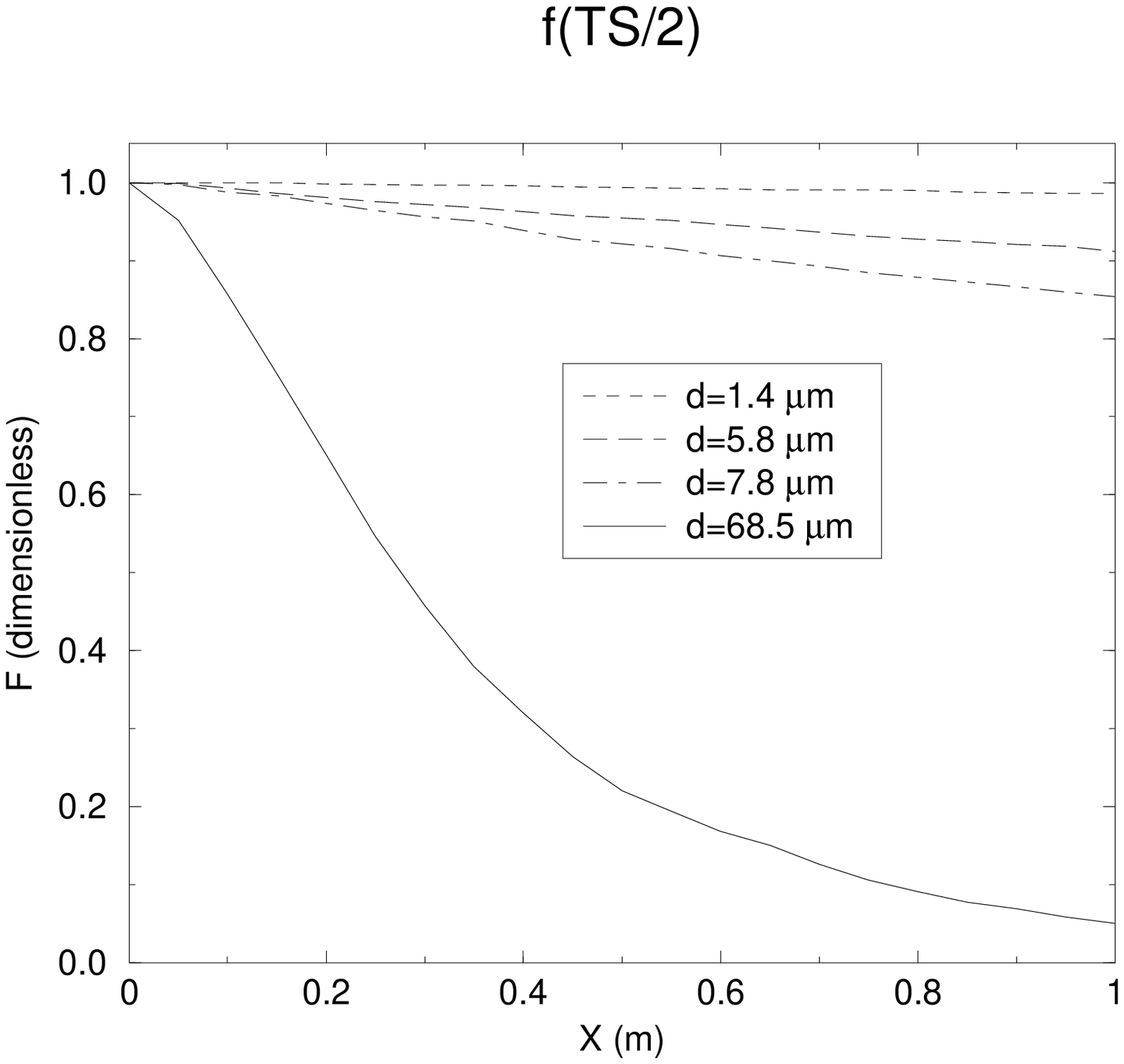,height=7.truecm,width=7.5truecm}
\caption{Fraction of particles remaining airborne versus pipe length for different diameter classes. These results are obtained with the phenomenological model, as explained in fig. \ref{bc}. Particles can deposit only after having stayed a certain residence-time in a small near-wall layer. The residence-time is different for each class of diameter and is computed for all classes through an empirical function f(Ts) deducted from DNS data, which give the correct value only for two classes. In Fig (a),  the results obtained using this function f(TS) derived from DNS data are shown. In Fig (b), results are shown the value of residence time given by f(TS) are divided by a factor two. It is seen that there is a little difference for the larger particles but that the difference is significant for the smaller. This indicates that the residence-time value used is crucial mainly for small particles. }
\label{F12}
\end{center}
\end{figure}

In Fig. \ref{F12}, we present the curve representing 
the fraction of particles remaining in the flow versus pipe axis
for the two functions of the characteristic time scale used, that is $T_s$
and $T_s/2$. 
For reasons of clarity, in the figure we show only
4 classes of diameters, which, however, represent all the regimes.
The figure shows again that, for small and intermediate diameters,
there is a noticeable difference in the fraction of particles which
deposit, while for large particles the behavior is very similar.

\begin{figure}[!h]
\begin{center}
\epsfig{file=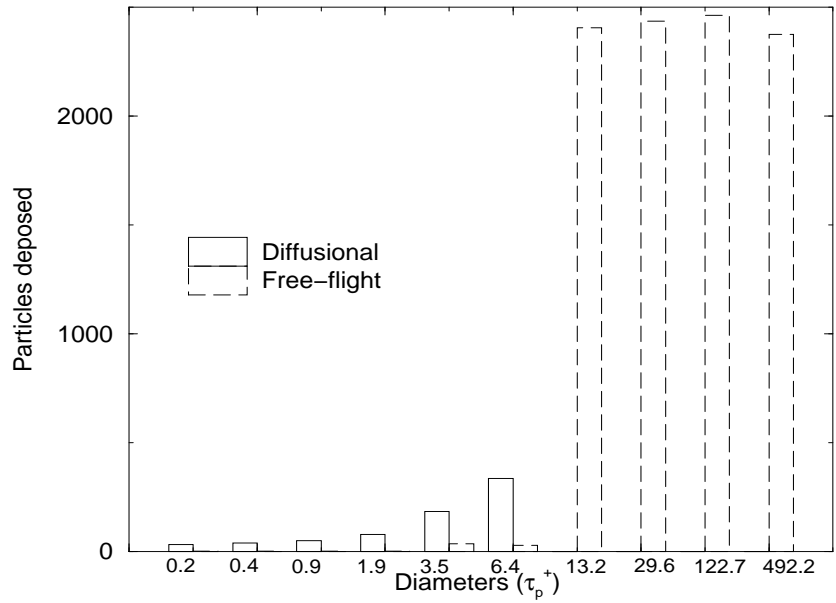,height=8.truecm,width=11.truecm}
\caption{Number of particles deposited for class of diameters and mechanism of deposition.}
\label{isto}
\end{center}
\end{figure}

In order to further assess the model function given by Eq. (\ref{dep_mod2})
we show, in Fig. \ref{isto},  the number of particles 
which deposit for each class of diameters,
in the case of the  function $T_s$.
We computed the fraction of particles deposited by the free-flight mechanism,
and the fraction of particles deposing by the diffusional one.
We can see that the model reproduces reasonably well the physical behavior
proposed by DNS calculations.
The diffusional mechanism is the most important for
small particles ($\tau_p^+ < 10$), while for the other classes 
free-flight mechanism becomes the only efficient one.
Moreover, the proportion between the two mechanisms
is correctly given, at least for light particles.
For the class of diameter $\tau_p^+ = 6.4$,
$80 \%$ of particles are found to deposit by diffusional mechanism,
while DNS results indicate a rate of $90 \%$ for $\tau_p^+ = 5$.
Finally, it is worth noting that the Lagrangian approach proposed in this work 
is grid-independent and valid for nominally infinite Reynolds-number,
thus should be easily used in much more complex 
geometries and grids. 

\section{Conclusions}

In this paper we have presented a numerical study of particle deposition in a 
turbulent pipe flow using a Langevin PDF model recently proposed~\cite{Min_04}.
In its standard formulation, the model has been found to be unable to reproduce 
the correct deposition velocity for light particles ($\tau^+_p<10$). Indeed,
results obtained with the standard form of the model have revealed a deposition 
velocity which is only slightly sensitive to particle inertia. Moreover, 
sensitivity tests have indicated that this outcome is not noticeably modified when mean fluid 
profiles are changed. As a simple and first step toward considering the specific effects
of coherent structures, a new phenomenological model, built on the basis of DNS results,
has been proposed and introduced in the numerical simulations. The results obtained 
with this model are in good agreement with experiments
and show that:
\begin{enumerate}
\item[(i)] One way to improve significantly the statistical description of particle 
deposition is to take into account some geometrical features of the flow, through the
sweeps and ejections events. In particular, residence-time in the near-wall 
region~\cite{Nar_03} which reflects their influence can be considered as a relevant parameter.
\item[(ii)] This model represents already a first attempt to relate a statistical 
description of the flow with a geometrical one, because we have put (crudely) some 
geometrical features of the instantaneous flow (coherent structures) in the 
framework of a statistical flow description, such as the present Langevin model.
This first proposition opens the road for a more systematic introduction 
of geometrical features in statistical PDF approach, where coherent structures in 
wall-bounded flows could be introduced as new stochastic terms in the modeled equations.
Given present results, this more explicit and rigorous stochastic approach appears as a 
good candidate for the construction of particle deposition models based on physical principles.
This is the subject of current research and of new stochastic models~\cite{Gui_07,Gui_07b}.
\item[(iii)] Even in its present formulation, the Langevin model proposed here yields 
satisfactory results and can be attractive for engineering applications, given its 
simplicity and stability (large time-steps can be used in the whole domain and for the 
whole range of particle diameter).
\end{enumerate}

\section{Acknowledgments}
S. Chibbaro's work is supported by a ERG EU grant.
This work makes use of results produced by the PI2S2 Project managed by the Consorzio COMETA, a project co-funded by the Italian Ministry of University and Research (MIUR). More information is available at http://www.pi2s2.it and http://www.consorzio-cometa.it. he greatly acknowledges the financial support given also by the consortium SCIRE.
We greatly acknowledge the important help of M. Ouraou for the fluid numerical computations. 
Moreover, S. Chibbaro would like to express special thanks to M. Ouraou for his constant,
patient
 and friendly support in all computational aspects during his stay at EDF.

\newpage
\begin{table}[htb]
\begin{center}
\begin{tabular}{|l|c|}
$\tau^+_p$  & Diameters ($\mu m$)  \\
\hline
$0.2$  & $1.4$  \\
\hline

$0.4$  & $2.0$ \\
\hline
$0.9$  & $2.9$   \\
\hline
$1.9$  & $4.3$      \\
\hline
$3.5$  & $5.8$      \\
\hline
$6.4$  & $7.8$    \\
\hline
$13.2$  & $11.2$ \\
\hline
$29.6$ & $16.8$     \\
\hline
$122.7$  & $34.2$   \\
\hline
$492.2$  & $68.5$   \\
\end{tabular}
\caption{Relation between $\tau^+_p$ and particle diameters.}
\label{tab:taup_dam}
\end{center}
\end{table}

\begin{table}[htb]
\begin{center}
\begin{tabular}{|l|c|c|}
$\tau^+_p$  & Diameters ($\mu m$) & Residence time (wall units) \\
\hline
$0.2$  & $1.4$ & 29.5 \\
\hline

$0.4$  & $2.0$ & 29.9 \\
\hline
$0.9$  & $2.9$ & 28.7  \\
\hline
$1.9$  & $4.3$ & 30.9     \\
\hline
$3.5$  & $5.8$ & 31.5     \\
\hline
$6.4$  & $7.8$ & 31.6   \\
\hline
$13.2$  & $11.2$ & 35.4  \\
\hline
$29.6$ & $16.8$  & 40.6   \\
\hline
$122.7$  & $34.2$  & 55.3 \\
\hline
$492.2$  & $68.5$  & 96.8 \\
\end{tabular}
\caption{Mean residence time for different classes of particle diameter in the case of 
exact mean fluid profiles, see $k^+-\epsilon^+$ curve in Fig.~\ref{viscous}.}
\label{sej1}
\end{center}
\end{table}

\listoffigures

\newpage

\newpage

\newpage

\newpage


\newpage

\newpage

\newpage

\end{document}